# Electronic structure and chemical bonding in Ti$_4$SiC$_3$ investigated by soft x-ray emission spectroscopy and first principle theory

M. Magnuson[1], M. Mattesini[1,4], O. Wilhelmsson[2], J. Emmerlich[3], J.-P. Palmquist[2], S. Li[1], R. Ahuja[1], L. Hultman[3], O. Eriksson[1] and U. Jansson[2]

[1]*Department of Physics, Uppsala University, P. O. Box 530, S-751 21 Uppsala, Sweden.*

[2]*Department of Materials Chemistry, The Ångström Laboratory, Uppsala University, P.O. Box 538 SE-75121 Uppsala.*

[3]*Department of Physics, IFM, Thin Film Physics Division, Linköping University, SE-58183 Linköping, Sweden.*

[4]*Departamento de Física de la Tierra, Astronomía y Astrofísica I, Universidad Complutense de Madrid, E-28040, Spain*

## Abstract

The electronic structure in the new transition metal carbide Ti$_4$SiC$_3$ has been investigated by bulk-sensitive soft x-ray emission spectroscopy and compared to the well-studied Ti$_3$SiC$_2$ and TiC systems. The measured high-resolution Ti *L*, C *K* and Si *L* x-ray emission spectra are discussed with *ab initio* calculations based on density-functional theory including core-to-valence dipole matrix elements. The detailed investigations of the Ti-C and Ti-Si chemical bonds provide increased understanding of the physical properties of these nanolaminates. A strongly modified spectral shape is detected for the buried Si monolayers due to Si 3*p* hybridization with the Ti 3*d* orbitals. As a result of relaxation of the crystal structure and the charge-transfer from Ti (and Si) to C, the strength of the Ti-C covalent bond is increased. The differences between the electronic and crystal structures of Ti$_4$SiC$_3$ and Ti$_3$SiC$_2$ are discussed in relation to the number of Si layers per Ti layer in the two systems and the corresponding change of materials properties.

## 1 Introduction

Ternary carbides and nitrides, also referred to as MAX-phases, denoted M$_{n+1}$AX$_n$, where n=1, 2 and 3 that we will refer to as 211, 312 and 413, respectively, have recently been the subject to intense research [1, 2, 3]. Here, M denotes an early transition metal, A is a p-element, usually belonging to the groups IIIA and IVA, and X is either carbon or nitrogen [4]. These nanolaminated-layered materials exhibit a unique combination of metallic and ceramic properties, including high strength and stiffness at high temperatures, resistance to oxidation and thermal shock, as well as high electrical and thermal conductivity [5]. The macroscopic properties are closely related to the underlying electronic structure and the structural properties of the constituent atomic layers. The MAX-phase family of compounds (over 50 variants are energetically stable) has a hexagonal structure with near close-packed layers of the M-elements interleaved with square-planar slabs of pure A-elements, where the X-atoms are filling the octahedral sites between the M-atoms. The A-elements are located at the center of trigonal prisms that are larger than the octahedral X sites. The structural difference between the 211, 312 and 413 phases is the number of inserted A-monolayers per M layer. The A/M ratios are 0.5, 0.33 and 0.25 for the 211, 312 and 413 phases, respectively. In addition, the 312 and 413 phases have two different M sites, denoted $M_I$ and $M_{II}$. The 413 crystal structure also has two different X-sites, denoted $X_I$ and $X_{II}$ and has more carbide-like attributes than the 211 and 312 crystal structures. The Ti-Si-C system in the MAX-phase family of compounds has been well studied, in particular the 312 type of crystal structure, which can be made both as thin film, and as sintered bulk material. $Ti_4SiC_3$ is the second known 413 compound after $Ti_4AlN_3$ and has only been synthesized as thin film [6, 7]. Therefore, the electronic structure of $Ti_4SiC_3$ is not as well known as the other MAX-phases and has more carbide like properties and less metallicity. Insertion of Si monolayers into a TiC matrix implies that the strong Ti-C bonds are replaced by weaker Ti-Si bonds. Thus, in $Ti_4SiC_3$, single monolayers of C atoms have been replaced by Si layers. The TiC layers surrounding the Si monolayers are then twinned with the Si layer as a mirror plane. Figure 1 shows the crystal structure of $Ti_4SiC_3$ with the nanolaminates of binary Ti-C-Ti slabs separated by softer Ti-Si-Ti slabs with weaker bonds. The elastic properties such as Young's modulus (E) change with phase and composition i.e., $Ti_4SiC_3$ is expected to be harder than the prototype compound $Ti_3SiC_2$ (320 GPa) which is softer than TiC (350-400 GPa) [8]. The change of elastic properties with phase is mainly due to the fact that the 413 crystal structure contains a larger fraction of strong Ti-C bonds compared to the 312 phases. The weaker Ti-Si bonds may also affect the tribological properties such as wear performance and friction. The physical properties of crystallographically oriented thin films of MAX phases can thus be custom made for a particular application such as protective coatings, sliding/gliding electrical contacts and heating elements.

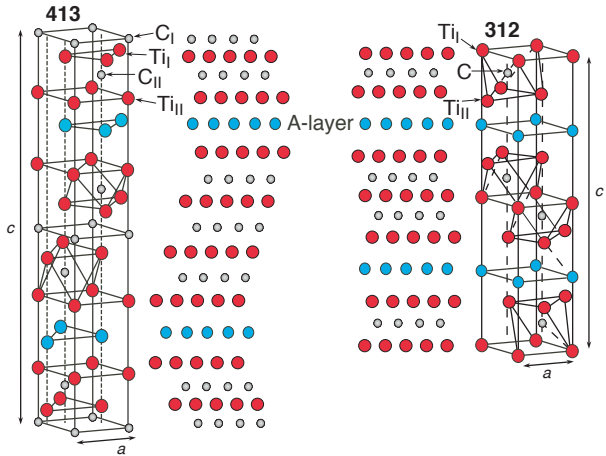

**Figure 1:** (Color online) The hexagonal crystal structure of 413 ($Ti_4SiC_3$) in comparison to 312. There is one Si layer for every fourth layer of Ti in $Ti_4SiC_3$. The lengths of the measured (calculated) $a$ and $c$-axis of the unit cell of $Ti_4SiC_3$ are 3.05 (3.08) Å and 22.67 (22.62) Å, respectively. Note that $Ti_I$ is connected to $C_I$ and $C_{II}$ while $Ti_{II}$ is connected to both $C_{II}$ and Si. $C_I$ is connected to $Ti_I$ type of atoms and $C_{II}$ is connected to both $Ti_I$ and $Ti_{II}$.

Theoretically, it has been shown by *ab initio* electronic structure calculations that there should be significant differences of the partial density-of-states (pDOS) of Ti, C and Si between different MAX-phase crystal structures [9, 10, 11, 12]. In recent studies, we investigated the three 312 phases $Ti_3AlC_2$, $Ti_3SiC_2$, $Ti_3GeC_2$ [13] and the 211 phase $Ti_2AlC$ [14]. In contrast to $Ti_3SiC_2$ and $Ti_3GeC_2$, a pronounced shoulder at 1 eV below the Fermi level ($E_F$) was identified in the Ti $L_{2,3}$ soft x-ray emission (SXE) spectra of $Ti_3AlC_2$ and $Ti_2AlC$. From these studies, it was clear that the physical and mechanical properties of MAX-phases can be further understood from detailed investigations of the electronic structure, in particular, the M-A and M-X chemical bonding.

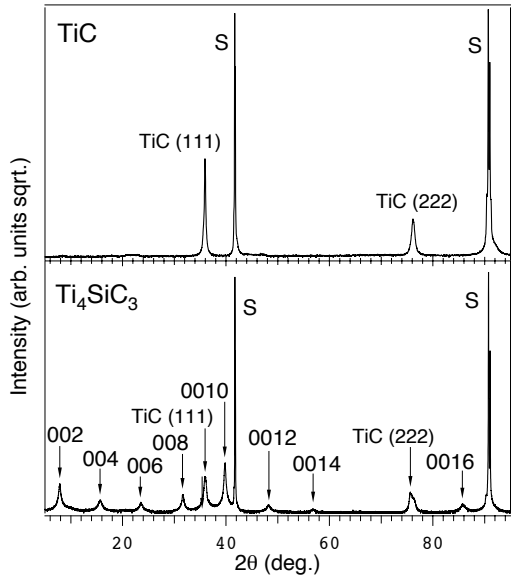

**Figure 2:** Top, x-ray diffractogram of TiC. Bottom, x-ray diffractogram from the $Ti_4SiC_3$ sample. S denotes the contribution from the $Al_2O_3$ substrate. The TiC peaks in $Ti_4SiC_3$ originates from the seed layer interface.

In the present paper, we investigate the electronic structure of $Ti_4SiC_3$, using bulk-sensitive and element-specific SXE spectroscopy with selective excitation energies around the Ti $2p$, C $1s$ and Si $2p$ thresholds. The SXE technique is more bulk sensitive than electron-based spectroscopic techniques such as x-ray absorption spectroscopy (XAS) and x-ray photoemission spectroscopy (XPS). Due to the involvement of both valence and core levels, the corresponding difference in energies of emission lines and their selection rules, each kind of atomic element can be probed separately. This makes it possible to extract both elemental and chemical near ground-state information of the electronic structure. The SXE spectra are interpreted in terms of partial valence band pDOS weighted by the transition matrix elements. The main objective of the present investigation is to study the nanolaminated internal electronic structures and the influence of hybridization among the constituent atomic planes in $Ti_4SiC_3$ in comparison to $Ti_3SiC_2$ and TiC, with the aim to obtain an increased understanding of the physical and mechanical properties.

## 2 Experimental

### 2.1 X-ray absorption and emission measurements

The SXE and XAS measurements were performed at the undulator beamline I511-3 at MAX II (MAX-lab National Laboratory, Lund University, Sweden), comprising a 49-pole undulator and a modified SX-700 plane grating monochromator [15]. The XAS spectra at the Ti $2p$ and C $1s$ edges were measured with 0.1 eV resolution. The SXE spectra were recorded with a high-resolution Rowland-mount grazing-incidence grating spectrometer [16] with a two-dimensional detector. The Ti $L$ and C $K$ SXE spectra were recorded using a spherical grating with 1200 lines/mm of 5 m radius in the first order of diffraction. The Si $L$ spectrum was recorded using a grating with 300 lines/mm, 3 m radius in the first order of diffraction. During the SXE measurements at the Ti $2p$, C $1s$, and Si $2p$ edges, the resolutions of the beamline monochromator were 1.6, 1.0 and 0.3 eV, respectively. The SXE spectra were recorded with spectrometer resolutions 0.7, 0.2, 0.2 eV, respectively. All the measurements were performed with a base pressure lower than $5\times10^{-9}$ Torr. In order to minimize self-absorption effects [17], the angle of incidence was about $20^o$ from the surface plane during the emission measurements. The x-ray photons were detected parallel to the polarization vector of the incoming beam in order to minimize elastic scattering.

## 2.2 Deposition of the Ti$_4$SiC$_3$ film

Figure 2 shows Θ-2Θ diffractograms of the deposited TiC and Ti$_4$SiC$_3$ films. The TiC$_x$(111) ($x$~0.7, 2000 Å thick) and Ti$_4$SiC$_3$ (000$l$) (900 Å thick) films were epitaxially grown on α-Al$_2$O$_3$ (000$l$) substrates at 300 $^o$C and 1000 $^o$C, respectively, by dc magnetron sputtering [18]. Elemental targets of Ti, C and Si, and a 3.0 mTorr Ar discharge were used. To promote a high quality growth of the MAX phase, a 250 Å thick seed layer of TiC$_{0.7}$(111) was initially deposited. For further details on the synthesis process, the reader is referred to Refs [19, 20, 21].

The two most intense peaks in Fig. 2 originate from the α-Al$_2$O$_3$(000$l$) substrate. As observed, the other peaks mainly originate from Ti$_4$SiC$_3$ (000$l$) together with peaks from the TiC(111) and (222) seed-layer. This strongly indicates a single-phase MAX material. Furthermore, the fact that the diffractogram shows only Ti$_4$SiC$_3$ of {000$l$}-type suggests highly textured or epitaxial films. X-ray pole figures verified that the growth indeed was epitaxial, and determined the relation to Ti$_4$SiC$_3$ (000$l$)//TiC(111)//Si$_2$O$_3$(000$l$) with an in-plane orientation of Ti$_4$SiC$_3$[210]//TiC[110]//Si$_2$O$_3$[210].

The values of the *a*-axis and *c*-axis were determined to be 3.05 and 22.67 Å by reciprocal space mapping (RSM). The epitaxial growth behavior has also been documented by transmission electron microscopy [22, 23, 24, 25, 26]. XPS-analysis depth profiles of the deposited films within the present study using a PHI Quantum instrument, showed after 60 seconds of Ar-sputtering a constant composition without any contamination species.

## 3 Computational details

### 3.1 Calculation of the x-ray emission spectra

The x-ray emission spectra were calculated within the single-particle transition model by using the APW+lo electronic structure method [27]. Exchange and correlation effects were described by means of the generalized gradient approximation (GGA) as parameterized by Perdew, Burke and Ernzerhof [28]. A plane wave cut-off, corresponding to $R_{MT}*K_{max}$=8, was used in the present investigation. For Ti and Si, *s* and *p* local orbitals were added to the APW basis set to improve the convergence of the wave function, while for C, only *s* local orbitals were added to the basis set. The charge density and potentials were expanded up to $\ell$=12 inside the atomic spheres, and the total energy was converged with respect to the Brillouin zone integration.

The SXE spectra were then evaluated at the converged ground-state density by multiplying the angular momentum projected density of states by a transition-matrix element [29]. The electric-dipole approximation was employed so that only the transitions between the core states with orbital angular momentum $\ell$ to the $\ell\pm1$ components of the electronic bands were considered. The core-hole lifetimes used in the calculations were 0.73 eV, 0.27 eV and 0.45 eV for the Ti 2$p$, C 1$s$ and Si 2$p$ edges, respectively. A direct comparison of the calculated spectra with the measured data was finally achieved by including the instrumental broadening in form of Gaussian functions corresponding to the experimental resolutions (see experimental section IIA). The final state lifetime broadening was accounted for by a convolution with an energy-dependent Lorentzian function with a broadening increasing linearly with the distance from the Fermi level according to the function $a+b$(E-E$_F$), where the constants *a* and *b* were set to 0.01 eV and 0.05 (dimensionless) [30].

## 3.2 Balanced crystal orbital overlap population (BCOOP)

In order to study the chemical bonding of the Ti$_4$SiC$_3$ compound, we calculated the BCOOP function by using the full potential linear muffin-tin orbital (FPLMTO) method [31]. In these calculations, the muffin-tin radii were kept as large as possible without overlapping one another (Ti=2.3 a.u., Si=2.3 a.u and C=1.6 a.u.), so that the muffin-tin radii fill about 66% of the total volume. To ensure a well-converged basis set, a double basis with a total of four different $\kappa^2$ values were used. For Ti, we included the 4$s$, 4$p$ and 3$d$ as valence states. To reduce the core leakage at the sphere boundary, we also treated the 3$s$ and 3$p$ core states as semi-core states. For Si, 3$s$, 3$p$ and 3$d$ states were taken as valence states. The resulting basis forms a single, fully hybridizing basis set. This approach has previously proven to give a well-converged basis [32]. For the sampling of the irreducible wedge of the Brillouin zone, a special-k-point method was used [33], and for the self-consistent total energy calculation, the number of k points were 216. In order to speed up the convergence, a Gaussian broadening of width 20 mRy was associated with each calculated eigenvalue.

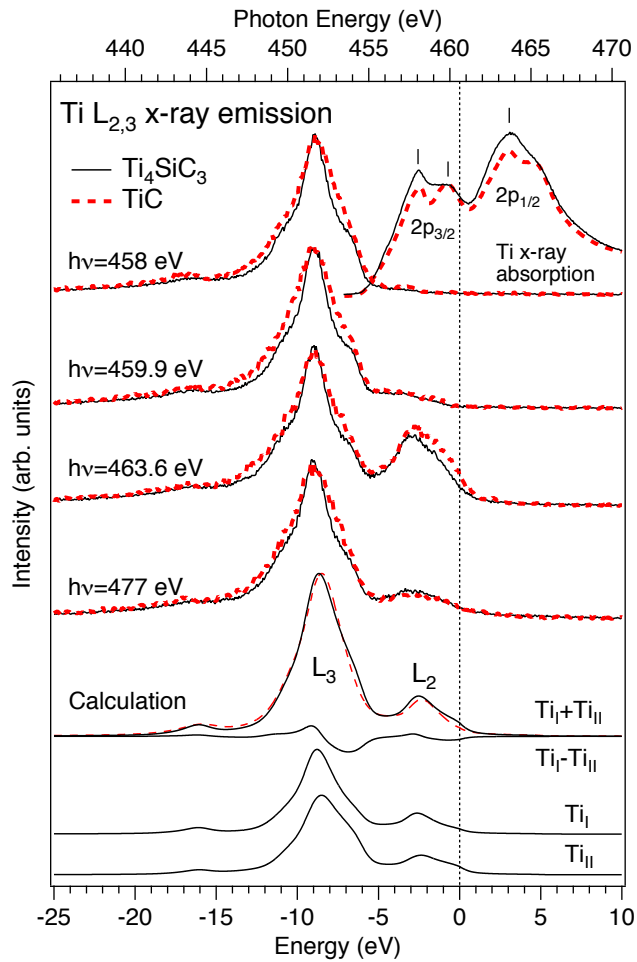

**Figure 3:** (Color online) Top, Ti $L_{2,3}$ x-ray emission spectra of Ti$_4$SiC$_3$ and TiC excited at 458, 459.9, 463.6 (resonant) and 477 eV (nonresonant). The excitation energies for the resonant emission spectra are indicated by vertical ticks in the x-ray absorption spectra (top, right curves). All spectra are aligned to the Ti $2p_{1/2}$ threshold at 460.6 eV measured by XPS on the Ti$_4$SiC$_3$ sample. Bottom, calculated spectra with fitted experimental $L_{2,3}$ peak splitting of 6.2 eV and the $L_3/L_2$ ratio of 6:1 compared to the x-ray emission spectra.

# 4 Results

## 4.1 Ti $L_{2,3}$ x-ray emission

Figure 3 (top) shows Ti $L_{2,3}$ SXE spectra of $Ti_4SiC_3$ excited at 458, 459.9, 463.6 and 477 eV photon energies, corresponding to the $2p_{3/2}$ and $2p_{1/2}$ absorption maxima and nonresonant excitation, respectively. The x-ray absorption measurements (top, right curves) were used to locate the excitation energies for the emission measurements. For comparison of the spectral shapes, the measured spectra are normalized to unity and are plotted on a photon energy scale (top) and a common energy scale (bottom) with respect to the $E_F$ using the Ti $2p_{1/2}$ core-level photoemission binding energy of 460.6 eV. The main $L_3$ and $L_2$ emission lines are observed at 2.5 eV and 9 eV on the common energy scale. Note that the Ti $L_{2,3}$ SXE spectral shapes of $Ti_4SiC_3$ and TiC are similar, indicating carbide-like attributes, although the main peak is somewhat broader in TiC. The energy dependence of the spectral shapes is rather weak with exception of the $L_2$ emission line which resonates at 463.6 eV, corresponding to the $2p_{1/2}$ absorption maximum. Weak peak features are observed around 16.5 eV in the measured spectra. A corresponding band feature is also observed in the calculated spectra at the bottom. The Ti $L_{2,3}$ SXE spectra are rather delocalized (wide bands) which makes electronic structure calculations suitable for the interpretation of the spectra. The fitted $L_3/L_2$ ratio was set to 6:1 as in the experimental spectra excited at 475 eV. The calculated spectra at the bottom are generally in good agreement with the experiment.

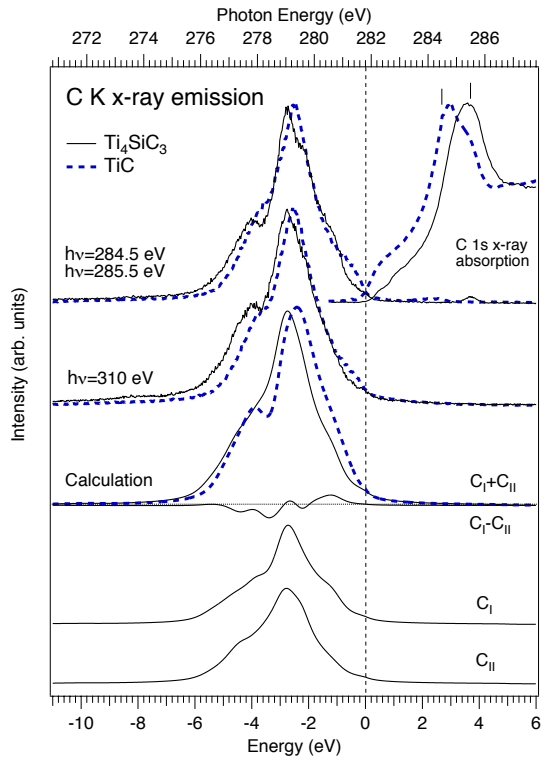

**Figure 4**: (Color online) Top, experimental C $K$ SXE spectra of $Ti_4SiC_3$ and TiC excited at 284.5, 285.5 (resonant) and 310 eV (non-resonant), aligned with the measured C 1$s$ core XPS binding energy 281.8 eV for $Ti_4SiC_3$. The resonant excitation energies for the SXE spectra are indicated in the C 1s XAS spectra (top, right curves) by the vertical ticks. Note the weak elastic peak at 285.5 eV in the resonant emission spectrum for $Ti_4SiC_3$. Bottom, calculated SXE spectra of $Ti_4SiC_3$ and TiC. The vertical dotted line indicates the Fermi level ($E_F$).

## 4.2 C *K* x-ray emission

Figure 4 (top) shows experimental C *K* SXE spectra of $Ti_4SiC_3$, excited at 284.5, 285.5 eV (resonant) and 310 eV (nonresonant) photon energies, respectively. The XAS spectra (top, right curves) were used to locate the excitation energies for the SXE spectra. Calculated SXE spectra are shown at the bottom. The main peak at 2.9 eV below $E_F$ is sharper at resonant excitation and has a pronounced shoulder on the low-energy side at 4.0 eV. Contrary to $Ti_3SiC_2$, $Ti_4SiC_3$, has no high-energy shoulder at 2 eV [13]. The agreement between the experimental and calculated spectra is generally good although the low-energy shoulder at 4.0 eV is less pronounced in the calculation. The main peak and the low-energy shoulder correspond to the occupied C 2*p* orbitals hybridized with the Ti and Si bonding and antibonding orbitals of the valence bands.

## 4.3 Si *L_{2,3}* x-ray emission

Figure 5 (top) shows experimental Si $L_{2,3}$ SXE spectra of $Ti_4SiC_3$, $Ti_3SiC_2$ and crystalline Si measured nonresonantly at 120 eV photon energy. Calculated SXE spectra are shown at the bottom. Comparing the experimental and calculated spectra, it is clear that the main peak at 7 eV of the SXE spectra is dominated by 3*s* final states. The partly populated 3*d* states form the broad peak structure closer to the $E_F$ and participate in the Ti 3*d* - Si 3*p* bonding in $Ti_4SiC_3$. Notably, the Si $L_{2,3}$ SXE spectrum of $Ti_4SiC_3$ has fewer sub-structures than $Ti_3SiC_2$ in the region 0-5 eV below $E_F$ [13]. This is an indication that the Ti 3*d* states of $Ti_4SiC_3$ hybridize different with Si than in $Ti_3SiC_2$ in this energy region. The difference in the Ti 3*d* - Si hybridization and chemical bonding is also attributed by the higher Si $2p_{1/2}$ XPS binding energy of $Ti_4SiC_3$ (100.0 eV) in comparison to $Ti_3SiC_2$ (98.5 eV) and Si(100) (99.5 eV). The Si 3*p* states dominate in the upper part of the Si $L_{2,3}$ spectrum but do not contribute to the spectral shape since they are dipole forbidden. For the Si $L_{2,3}$ SXE spectrum, the valence-to-core matrix elements are found to play an important role to the spectral shape. In contrast to the Si $L_{2,3}$ SXE spectrum of pure Si, which has a pronounced double structure, the Si $L_{2,3}$ SXE spectra of $Ti_4SiC_3$ and $Ti_3SiC_2$ have strongly modified spectral weights towards the $E_F$. A similar modification of the Si $L_{2,3}$ SXE spectral shape has also been observed in the metal silicides [34]. Comparing the Si $L_{2,3}$ SXE spectral shapes of $Ti_4SiC_3$ and $Ti_3SiC_2$ to the silicides, the appearance of the shoulder around 9 eV can be attributed to the formation of hybridized Si 3s states produced by the overlap of the Ti 3d-orbitals. This interpretation is supported by our first principle calculations.

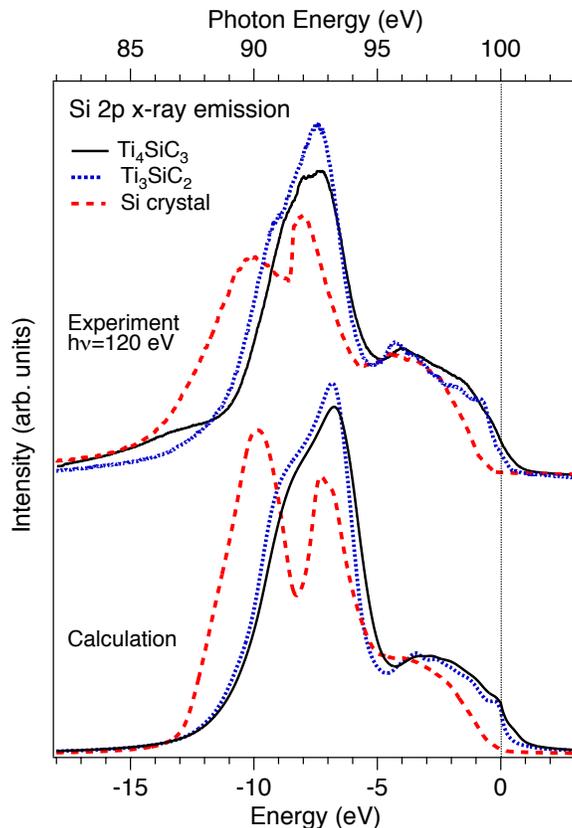

**Figure 5:** (Color online) Top, experimental Si $L_{2,3}$ spectra of $Ti_4SiC_3$ compared to $Ti_3SiC_2$ [13] and crystalline Si [100], all excited at 120 eV. Bottom, calculated spectra. The spectra were normalized to the total area and the vertical dotted line indicates the Fermi level ($E_F$).

## 4.4 Chemical Bonding

By relaxing the cell parameters of $Ti_4SiC_3$, it was possible to calculate the equilibrium $a$ and $c$-axis. They were determined to be 3.08 and 22.62 Å for $Ti_4SiC_3$. These values are in good agreement with the experimental values of 3.05 and 22.67 Å in section IIB. In order to analyze the chemical bonding in more detail, we show in figure 6 the calculated BCOOP [35] of the $Ti_4SiC_3$ system compared to $Ti_3SiC_2$ [13] and TiC. The BCOOP makes it possible to compare the strength of two similar chemical bonds. The BCOOP is a function, which is positive for bonding states, and is negative for anti-bonding states. The strength of the covalent bonding can be determined by summing up the area under the BCOOP curve. The energy position of the peaks also gives an indication of the strength of the covalent bonding. Firstly, comparing the areas under the BCOOP curves and the distances of the main peaks of the curves from the Fermi level, it is clear that the Ti 3d - C 2p bonds are much stronger than the Ti 3d - Si spd bonds in both $Ti_4SiC_3$ and $Ti_3SiC_2$. The Ti atoms loose some bond strength to the nearest neighbor Si atoms, which to some degree is compensated with a stronger Ti-C bond. Secondly, comparing the BCOOP curves of $Ti_4SiC_3$ to those of $Ti_3SiC_2$, the Ti-C BCOOP curve of $Ti_4SiC_3$ is less intense which indicates that the Ti-C bond is somewhat weaker in $Ti_4SiC_3$ than in $Ti_3SiC_2$. It should be noticed that the $Ti_{II}$-$C_{II}$ bonds are shorter than the Ti-C bonds in TiC (see Table I). This implies that the bonds in the Ti-C slabs of the MAX-phase are stronger than in TiC and is due to the weaker Ti-Si bonds, which transfer charge to the Ti-C bonds.

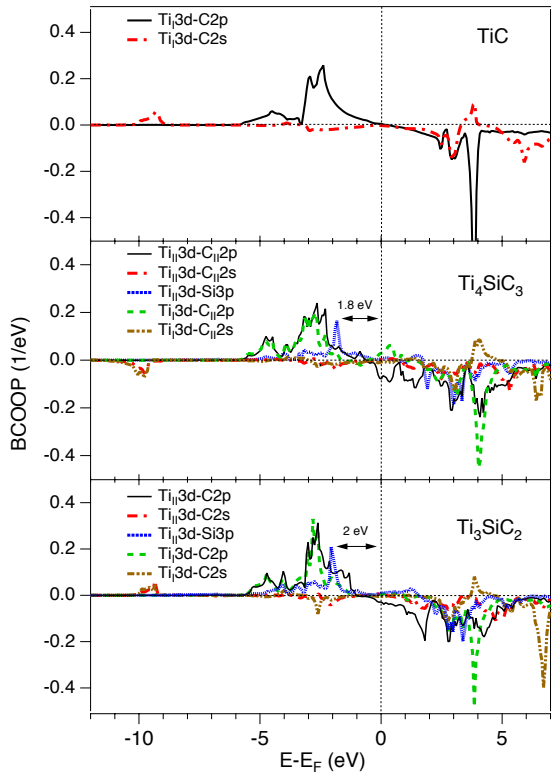

**Figure 6:** (Color online) Calculated balanced crystal overlap population (BCOOP) of TiC, $Ti_4SiC_3$ and $Ti_3SiC_2$. In $Ti_4SiC_3$, $Ti_I$ is connected to $C_I$ and $C_{II}$ while $Ti_{II}$ is connected to both $C_{II}$ and Si. $C_I$ is connected to $Ti_I$ type of atoms and $C_{II}$ is connected to both $Ti_I$ and $Ti_{II}$ as illustrated in Fig. 1. Note that the Ti 3d and C 2s overlap around 10 eV below $E_F$ is antibonding in $Ti_4SiC_3$ and bonding for TiC and $Ti_3SiC_2$.

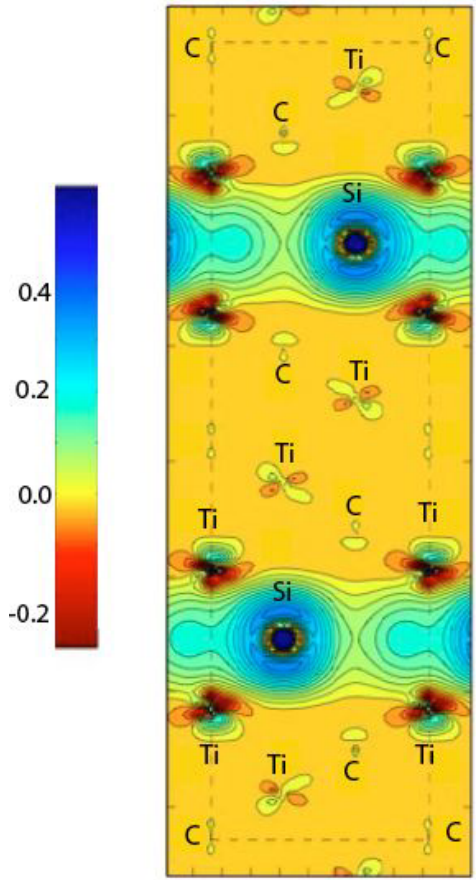

**Figure 7:** Calculated charge-density difference between $Ti_4SiC_3$ and $Ti_4C_4$ (TiC) in the same crystal geometry. A carbon atom is located at each corner of the plot where the charge-density difference is zero. The difference density plot was obtained by subtracting the charge densities in the [110] diagonal plane of the hexagonal unit cell. The lower valence band energy was fixed to -1.0 Ry (-13.6 eV) and all the Ti 3$d$, 4$s$; Si 3$s$, 3$p$ and C 2$s$, 2$p$ valence states were taken into account.

For the Ti $L_{2,3}$ SXE spectra of $Ti_4SiC_3$, discussed in section IVA, the BCOOP calculations confirm that the Ti 3$d$-C 2$p$ hybridization and strong covalent bonding is in fact the origin of the main peak at 9 eV below the $E_F$. Although there is a single Ti-C peak, the BCOOP analysis show that there are many overlapping energy levels in the energy region 0-6 eV below $E_F$. The Ti-Si BCOOP peak of $Ti_4SiC_3$ at 1.8 eV is less intense and closer to the $E_F$ than in $Ti_3SiC_2$ (2.0 eV). This is an indication that the Ti-Si chemical bond in $Ti_4SiC_3$ is weaker than in $Ti_3SiC_2$ and thus plays a key role for the physical properties. The states near $E_F$ are dominated by Ti 3$d$ orbitals with contribution from Si 3$p$ orbitals. However, there is also metal-metal $dd$ interactions (metal bonding) close to $E_F$. A strengthening of the Ti-Si $pd$ covalent bonding should in principle increase the shear stiffness (hardness and elasticity) although the most important mechanism is the number of Si layers inserted in the TiC matrix. The calculated C-Si overlaps (not shown) have a very different shape in comparison to the other overlaps, which is an indication that this bond has a non-covalent character.

Figure 7 shows a calculated electron density difference plot between $Ti_4SiC_3$ and $Ti_4C_4$, where in the latter system Si has been replaced by C in the same 413 crystal structure representing a highly twisted TiC structure. When introducing the Si atoms into the $Ti_4C_4$ crystal structure we first observe an anisotropic charge variation around the Ti atoms close to Si. In particular, in the direction along the Ti-Si bond (~ 45$o$ angle to the corners of the plot) we register an electron density withdrawal (see the red/dark area around Ti) from Ti to Si as to indicate the formation of the Ti-Si bonds. The consequence of such an electronic movement is the creation of a certain polarization on the neighbor Ti-Ti bonding. The insertion of the Si atoms in the $Ti_4C_4$ structure introduce an anisotropic electron density distribution primarily in a thin sheet containing Ti and Si atoms, resulting in a whole charge-modulation along the Ti-Si-Ti-Ti-Ti-Ti-Si-Ti zigzag bonding direction that propagates throughout the unit cell. Finally, we also observe that the charge-density difference is zero at the carbon atoms at the corners of the plot in Fig. 7. This is an indication that the carbon atoms do not respond markedly to the introduction of Si planes and implies that Si substitution only results in local modifications to the charge density, and possibly a weak Si-C interaction.

**Table 1:** Calculated bond lengths for $Ti_4SiC_3$ $Ti_3SiC_2$ and TiC. In $Ti_4SiC_3$, $Ti_I$ is connected to $C_I$ and $C_{II}$ while $Ti_{II}$ is connected to both $C_{II}$ and Si. $C_I$ is connected to the $Ti_I$ type of atoms and $C_{II}$ is connected to both $Ti_I$ and $Ti_{II}$ as illustrated in Fig. 1.

| Bond type | $Ti_I$-$C_I$ | $Ti_I$-$C_{II}$ | $Ti_{II}$-$C_{II}$ | $Ti_{II}$-$C_I$ | $Ti_{II}$-Si |
|---|---|---|---|---|---|
| TiC | 2.164 | - | - | - | - |
| $Ti_3SiC_2$ | - | 2.189 | 2.097 | - | 2.694 |
| $Ti_4SiC_3$ | 2.151 | 2.215 | 2.093 | 3.631 | 2.695 |

# 5 Discussion

Comparing $Ti_4SiC_3$ with $Ti_3SiC_2$, it is clear that the physical properties and the underlying electronic structure of the Ti-Si-C system is strongly affected by the number of Si layers per Ti layer. However, a large fraction of the charge-transfer indeed comes from the Si atoms resulting in a charge modulation along the c-axis. Our charge-density difference calculations for the $Ti_4SiC_3$ system along the [110] plane show that the C atoms have large gain of electron density whereas Ti and Si loose charge even though the choice of phase changes the Ti-Si chemical bond. All the Ti-Si-C MAX phases show excellent conductivity due to the metallic component of the bonding. Intuitively, one would therefore expect that the conductivity would decrease as more Si monolayers are introduced since Si is a semiconductor. However, in $Ti_4SiC_3$, the $E_F$ is close to a pronounced pseudogap (a region with low density of states). The conductivity is largely proportional to the number of states at the $E_F$ (TiC: 0.12 states/eV/atom, $Ti_3SiC_2$: 0.33 states/eV/atom, $Ti_4SiC_3$: 0.29 states/eV/atom [25]). The 413 system thus has a lower conductivity than the 312 system due to the decreasing metallicity in the pseudogap. In our previous 312 study [13], it was clear that the $Ti_{II}$ layers contribute more to the conductivity than the $Ti_I$ layers. However, the electrical and thermal conductivity is higher in the 312 and 211 systems. Comparing $Ti_4SiC_3$ with $Ti_3SiC_2$, one can anticipate that the E-modulus increases with decreasing number of Si layers per Ti layer. The hardening of the 413 phase (more carbide-like attributes) is due to changes in the bonding conditions of the weaker Ti-Si bonds. In this sense, $Ti_4SiC_3$ is more similar to TiC than $Ti_3SiC_2$ since there is a reduced number of inserted Si monolayers. Concerning the deformation and delamination mechanism, it is similar in all MAX-phases due to the weak M-A bonds. Our results show a clear difference between the electronic structures of the two MAX-phases $Ti_4SiC_3$ and $Ti_3SiC_2$ depending on the number of Si layers per Ti layer. The properties of the Ti-Si-C systems are thus directly related to the number of inserted Si layers into the TiC matrix. This is due to the much weaker covalent bond between Ti and Si compared to the strengthened Ti-C bond, which hardens and stiffens the material.

# 6 Conclusions

In summary, we have investigated the electronic structure and chemical bonding in $Ti_4SiC_3$ and compared the results to $Ti_3SiC_2$ and TiC with the combination of soft x-ray emission spectroscopy and electronic structure calculations. The covalent bonding mechanism is found to be very important for the physical properties such as hardness. The combination of experimental and theoretical results show that the Ti $3d$ - Si $3p$ bonding in $Ti_4SiC_3$ has a relatively weak covalent character. The calculated orbital overlaps also indicate that the Ti-Si bonding orbitals of $Ti_4SiC_3$ are somewhat weaker than in $Ti_3SiC_2$ which implies a change of the elastic properties and the electrical and thermal conductivity. The analysis of the underlying electronic structure thus provides increased understanding of the difference of materials properties between the $Ti_4SiC_3$ and $Ti_3SiC_2$ compounds. As in the case of $Ti_3SiC_2$, the x-ray emission spectra of Si in $Ti_4SiC_3$ appear very different from crystalline Si indicating strong hybridization between Si atoms and Ti with less influence from C. Tuning of the physical and mechanical properties implies that these nanolaminated carbide systems can be custom-made by the choice of crystal structure and the insertion of more or fewer Si layers in the TiC matrix.

# 7 Acknowledgements

We would like to thank the staff at MAX-lab for experimental support. This work was supported by the Swedish Research Council, the Göran Gustafsson Foundation, the Swedish Strategic Research Foundation (SSF) Materials Research Programs on Low-Temperature Thin Film Synthesis and the Swedish Agency for Innovation Systems (VINNOVA) Project on Industrialization of MAX Phase Coatings.